\documentclass[journal=nalefd,manuscript=letter]{achemso}
\usepackage[colorinlistoftodos]{todonotes}
\usepackage[version=3]{mhchem} % Formula subscripts using \ce{}
\usepackage{siunitx}
\usepackage{amsmath}
\usepackage{amssymb}
\author{Andrii Trelin}
\affiliation{Institute of Physics, University of Rostock, 18059 Rostock, Germany}
\alsoaffiliation{Department of Life, Light, and Matter of the Interdisciplinary Faculty at Rostock University, 18059 Rostock, Germany}

\author{Sophie Kussauer}
\affiliation{Department of Cardiac Surgery, Rostock University Medical Centre, 18057 Rostock, Germany}
\alsoaffiliation{Department of Life, Light, and Matter of the Interdisciplinary Faculty at Rostock University, 18059 Rostock, Germany}

\author{Paul Weinbrenner}
\affiliation{Institute of Physics, University of Rostock, 18059 Rostock, Germany}
\alsoaffiliation{Department of Life, Light, and Matter of the Interdisciplinary Faculty at Rostock University, 18059 Rostock, Germany}

\author{Anja Clasen}
\affiliation{Institute of Physics, University of Rostock, 18059 Rostock, Germany}
\alsoaffiliation{Department of Life, Light, and Matter of the Interdisciplinary Faculty at Rostock University, 18059 Rostock, Germany}

\author{Robert David}
\affiliation{Department of Cardiac Surgery, Rostock University Medical Centre, 18057 Rostock, Germany}
\alsoaffiliation{Department of Life, Light, and Matter of the Interdisciplinary Faculty at Rostock University, 18059 Rostock, Germany}

\author{Christian Rimmbach}
\affiliation{Department of Cardiac Surgery, Rostock University Medical Centre, 18057 Rostock, Germany}
\alsoaffiliation{Department of Life, Light, and Matter of the Interdisciplinary Faculty at Rostock University, 18059 Rostock, Germany}
\altaffiliation{These authors contributed equally}

\author{Friedemann Reinhard}
\affiliation{Institute of Physics, University of Rostock, 18059 Rostock, Germany}
\alsoaffiliation{Department of Life, Light, and Matter of the Interdisciplinary Faculty at Rostock University, 18059 Rostock, Germany}
\email{friedemann.reinhard@uni-rostock.de}
\altaffiliation{These authors contributed equally}

\title[An \textsf{achemso} demo]
  {%ChiSCAT: a chaotic microscopy setup and unsupervised learning algorithm for its analysis \\
  ChiSCAT: unsupervised learning of recurrent cellular micro-motion  patterns from a chaotic speckle pattern \\
    }


\begin{document}

\begin{tocentry}
\includegraphics[width=8.47cm]{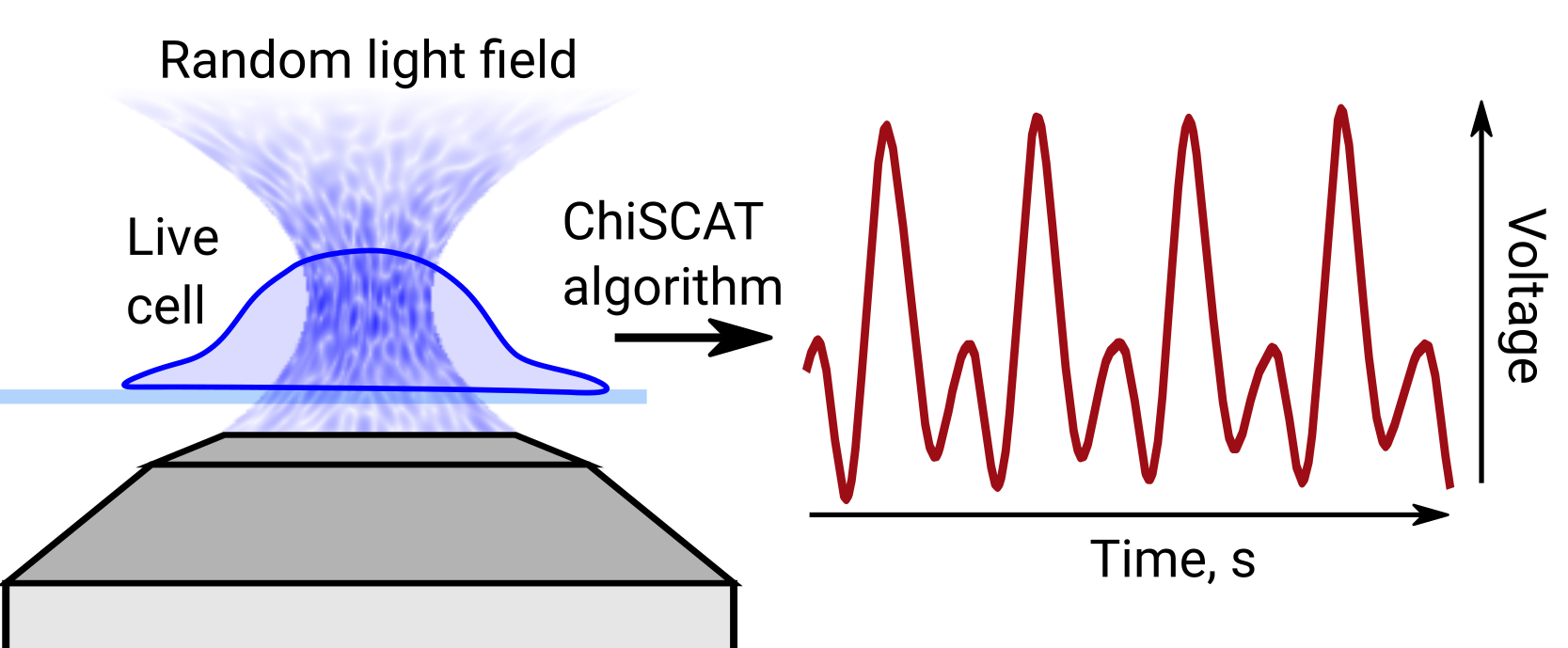}
\end{tocentry}

\begin{abstract}

% - Currents techniques are bad -> need something better. Technique could work in scattering tissue (brain), wavelength freely chosen -> image entire cortex 
% - turning iSCAT upside down -> maximize speckles, recover by analysis
% - Advantages:
%     - No techniques have been able to measure it reliably so far
%     - Reflective scheme -> sensitive to motions in the whole cell body + absolute motion of the cell
%     - could work in scattering tissue (brain)
%     - image entire cortex
%     - Insensitive to vibration
%     - minimal a priori knowledge (rough temporal shape + repetition) 

There is considerable evidence that action potentials are accompanied by "intrinsic optical signals", such as a nanometer-scale motion of the cell membrane.
Here we present ChiSCAT, a technically simple imaging scheme that detects such signals with interferometric sensitivity. ChiSCAT combines illumination by a {\bf ch}aotic speckle pattern and interferometric scattering microscopy ({\bf iSCAT}) to sensitively detect motion in any point and any direction. The technique features reflective high-NA illumination, common-path suppression of vibrations and a large field of view. 
This approach maximizes sensitivity to motion, but does not produce a visually interpretable image. We show that unsupervised learning based on matched filtering and motif discovery can recover underlying motion patterns and detect action potentials. 
%with an only moderate overhead in sensitivity over the fundamental limit achievable by linear filtering. 
We demonstrate these claims in an experiment on blebbistatin-paralyzed cardiomyocytes. ChiSCAT promises to even work in scattering tissue, including a living brain.
%{minimal a priori information - rough temporal template, }
\end{abstract}

\section{Introduction}
Literature data suggests that action potentials are accompanied by "intrinsic optical signals", tiny changes in the optical properties of a cell that can be detected by label-free optical imaging \cite{hill49, cohen68, iwasa80,  laporta1990recording,stepnoski1991noninvasive,macvicar1991imaging,kim2007mechanical,laporta2012interferometric,badreddine2016real, yang2018imaging, ling2018full,ling2020high}. The most accepted mechanism is a nanometer-scale motion of the cell membrane \cite{kim2007mechanical}, occurring due to various factors, including the flexoelectric effect or voltage-induced changes in membrane tension \cite{mueller2014quantitative}. Detecting action potentials by direct label-free observation of nanoscale cell motions in interferometric microscopy is highly promising, since it provides several unique advantages over dye-based voltage or calcium imaging: it does not suffer from photobleaching, which enables prolonged observation. Moreover, the illumination wavelength can be freely chosen and can be optimized to minimize cellular damage or to minimize optical scattering and hence 
allow chronical imaging through multiple tissue layers, e.g. the cortex of the brain.
\par 
Intrinsic signals are a small fluctuation on a large background. Past studies have established that the average scale of membrane movement is $\Delta z\approx 1$ nm \cite{mosbacher1998voltage,kim2007mechanical,gonzalez2016solitary,ling2018full,ling2020high}. In a suitable phase sensitive interferometric microscopy technique, this translates into an intensity change of 

$$  \frac{\Delta I_{\text{signal}}}{I} \approx \Delta \phi \approx \frac{\Delta n}{n} \frac{\Delta z}{\lambda} \approx 10^{-5}$$
 assuming a relative refractive index change between cell and medium ${\Delta n / n} \approx 10^{-2}$ (according to \cite{choi2007tomographic}) and a  relative phase change of $\Delta z / \lambda\approx \num {1e-3}$. 
 Intriguingly, such a signal should be easily detectable in a single shot. The quantum shot noise limit of interferometry is 
 $$
 \frac{\Delta I_{\text{shot noise}}}{I} = \sqrt{\frac{1}{N}}
 $$
 where $N=P\Delta t/(\hbar\omega)$ denotes the number of photons employed. Assuming $P=1$ mW of illumination power and an integration time of $\Delta t=1$ ms, matching the timescale of an action potential (AP), the quantum shot noise limit is on the order of $\Delta I_{\text{shot noise}}/{I}\approx 10^{-7}$, and the expected signal-to-noise ratio (SNR) is around 100.

%no single cell single shot detection to date
However, existing experiments on label-free AP detection still fall short of this limit. Published results either have been obtained on non-mammalian large axons and nerves \cite{laporta1990recording}, had to resort to invasive techniques like atomic force microscopy \cite{zhang01} or plasmon imaging \cite{howe2019surface}, or to spike-triggered averaging of multiple APs to reveal the signal with SNR $\geq 1$ \cite{ling2018full,ling2020high,yang2018imaging}. 

Presumably, a major reason is that increasing the SNR is more challenging than merely raising the excitation power. In transmission microscopy, the full-well capacity of the imaging camera sets a limit for excitation power at the level of nanowatts. This can be overcome by a reflective illumination scheme, promising a great gain in sensitivity \cite{ling2020high}. However, existing reflective schemes either restrict acquisition to a single diffraction-limited plane \cite{yaqoob11, singh19}, reducing the signal, or merely produce an unspecific speckle pattern, which cannot be easily interpreted \cite{popescu2019quantitative}: the reflected beam contains interfering signals from all depth levels inside the cell, which leads to chaotic speckle-like appearance. 

Here, we break this limitation. We present an unsupervised learning scheme to recover AP signals from a speckle pattern, without {\em a priori} knowledge of the timing of the action potentials or the spatial pattern of motion triggered during an AP. Moreover, we intentionally maximize the chaotic nature of the image by illuminating the specimen with a speckle pattern. This creates a random 3D light field in the cell, making the technique sensitive not just to the motion in height, but also to motion in X-Y plane, achieving interferometric sensitivity to motion in any illuminated voxel. In other words, the observed image is a "fingerprint" of the cell geometry and exact volume distribution of intracellular scatterers. 

Since ChiSCAT is not based on a visually interpretably image, it paves the way to detecting action potentials even in scattering tissue, including the  living brain: signal passing through multiple layers of scatterers will change, but it's nature of a cell shape "fingerprint" will be conserved. 

%The key advantage of reflective illumination is significant increase of relative contrast. While the term $\Delta n / n$ scales phase change in trasmissive scheme, it scales intensity in reflective scheme, which greatly simplifies signal detection with finite well capacity cameras.

%Probable explanation comes from technical problems, main difficulty of detection tiny signal changes on a large background (which will simply saturate the camera at 1 mW) and suppressing other noise sources (dominated by vibrations in interferometric setup).

%To achieve AP detection, experimental setup must satisfy following conditions:
%\begin{itemize}
%    \item Maximal achievable sensitivity to motion, which requires using high-NA illumination
%    \item High robustness to vibration, which is only achievable in common-path interferometric schemes
%    \item Signal-to-background ratio, which can be handled by existing cameras. This can be achieved by switching to reflection illumination  (explained below in details)
%    \item Sensitivity to in-plane motion, where motion amplitude is expected to be greater \cite{elhady2015mechanical}.

%\end{itemize}

\section{Setup}

\begin{figure}
  % As well as the standard float types \texttt{table}\\
  % and \texttt{figure}, the class also recognises\\
  % \texttt{scheme}, \texttt{chart} and \texttt{graph}.
  \includegraphics[width=1.0\textwidth]{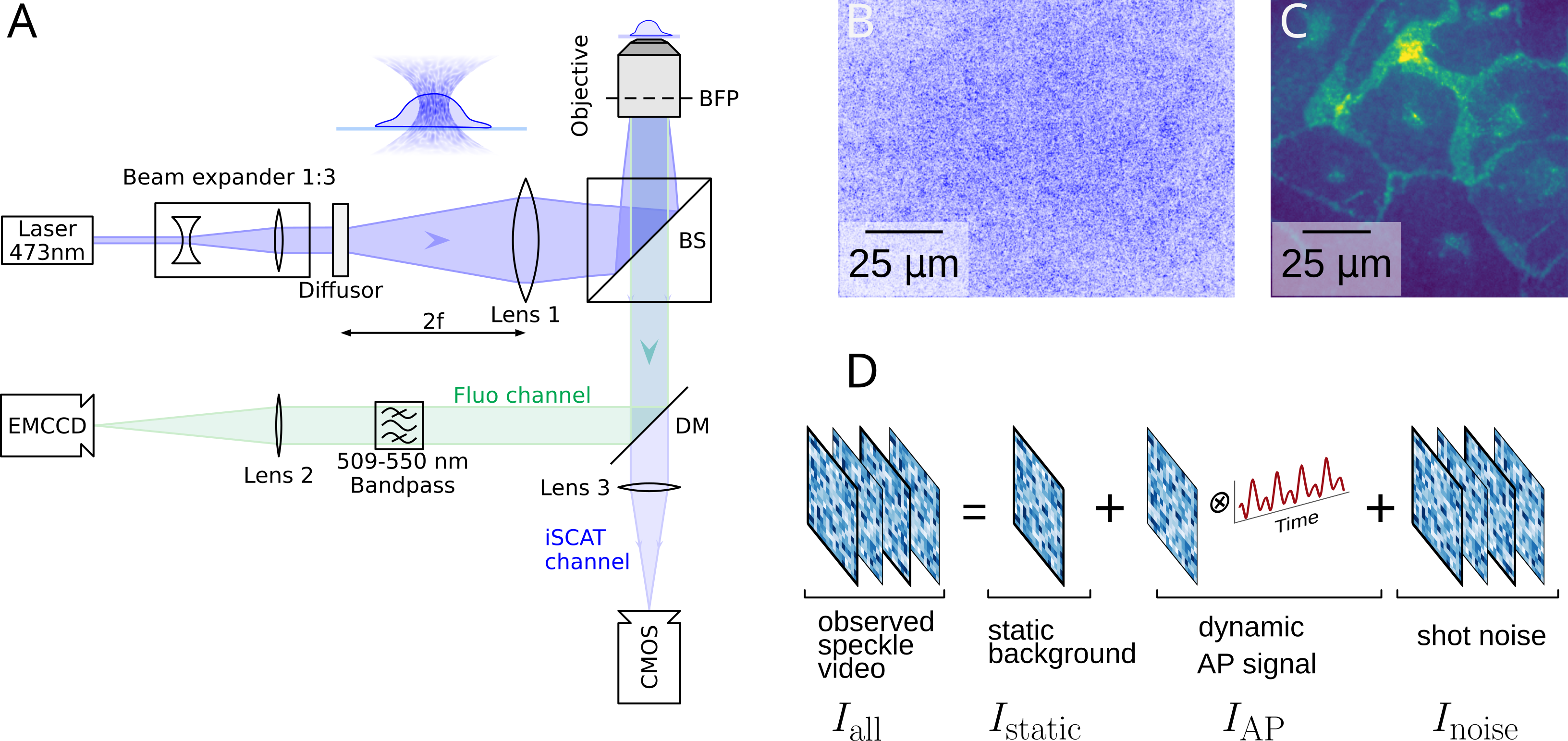}
  \caption{Experimental setup and exemplary data. (A) Optical setup, combining iSCAT microscopy with speckle illumination (ChiSCAT) and a fluorescence voltage imaging path for reference. Both cameras are synchronized. (B) Raw ChiSCAT image (C) raw fluorescence image (D) model of the ChiSCAT signal. A static speckle pattern background ("Static speckle") is superimposed with a periodically modulated speckle pattern ("AP speckle") which repeats every action potential and arises from cellular micromotion. }
  \label{fig:setup}
\end{figure}

Our setup (Fig.~\ref{fig:setup} A) is based on widefield interferometric scattering microscopy (iSCAT). We illuminate the cell with coherent light. The resulting image is an interference of light scattered by the cell and light scattered at the cover slide-immersion medium (water) interface, providing common-path suppression of vibrations in the setup. Light scattered in random parts of the cell interferes, forming a speckle signal. In iSCAT, this speckle background is usually carefully suppressed, e.g. to highlight motion of a single nanoparticle \cite{piliarik2014direct}. 
ChiSCAT takes an opposite approach. In order to generate a motion fingerprint of the entire cell, we maximize the speckle background by employing speckle illumination (Fig.~\ref{fig:setup}A+B). This is achieved by a holographic diffusor in the excitation path. Importantly, speckle illumination uses the full numerical aperture of the microscope. Both illumination and detection have a high numerical aperture, yielding a higher resolution than classical widefield iSCAT or quantitative phase imaging, which employ a parallel, i.e. low-NA, excitation beam. The setup features a correlative path for fluorescence voltage imaging (Fig.~\ref{fig:setup}A+C), which is only needed as a ground-truth reference to benchmark the processing algorithms. We employ this setup to perform experiments on  human induced pluripotent stem cells derived cardiomyocytes, with mechanical contraction has been disabled by the myosin-inhibitor blebbistatin. Since this inhibition might not be perfect, these cells can potentially  exhibit larger micromotion in response to an AP than neurons. Most likely, the motion is larger than few nanometers, but still significantly below the optical diffraction limit ($\approx$ 240 nm).

In the ChiSCAT channel we expect a signal as sketched in Figure~\ref{fig:setup} D. It consists of three components: a static speckle background created mainly by light scattered from the coverslide, a speckle pattern which is modulated by the time-dependent and repeating AP signal, and noise which for sufficiently strong illumination is dominated by photon shot noise.

Mathematically, the superposition of two of these three components in the camera plane can be modeled as a complex (phasor) electric fields
$$
E_{\text{all}} = E_{\text{static}} + E_{\text{AP}}
$$
Here, $E_{\text{static}}$ refers to light reflected from the cover slide and static parts of the cell, and $E_{\text{AP}}$ to light scattered from cellular components varying along the AP, e.g. a vibrating plasma membrane. 
All fields $E_{\text{all$|$static$|$AP}}(x,y,t)$ are random speckle fields in the spatial coordinates $x$ and $y$, and can depend on time ($t$). 
The intensity recorded by the camera is then given by 
$$
I_{\text{all}} = E_{\text{all}}^*E_{\text{all}} \approx I_{\text{static}} + I_{\text{AP}} + I_{\text{noise}}
$$
Here, a noise term $I_{\text{noise}}$ has been added. Crucially, the AP signal can be factorized as follows
\begin{gather*}
E_{\text{AP}}(x,y,t) = E_{\text{AP}}(x,y) \otimes a(t);\qquad \\
I_{\text{AP}}(x,y,t) = \text{Re}\left[E_{\text{static}}(x,y)^*\cdot E_{\text{AP}}(x,y)\right] \otimes a(t) =: I_{\text{AP}}(x,y)\otimes a(t);
\end{gather*}
where $a(t)$ denotes action potential amplitude, likely following a similar time trace as the trans-membrane voltage. In a simplified picture, whenever $a(t)$ spikes, a characteristic pattern $I_{\text{AP}}(x,y)$ appears in the ChiSCAT recording. 

% Methods

% The fluorescence channel was only needed as a source of ground truth for data processing algorithm development. Due to its low intensity, presence of dichroic mirror and high camera framerate, no fluorescence signal can be detected in the recorded speckle videos. The detected signal is a mixture of cell speckle (originating from interference of reflections from every cell voxel), weighted by the motion pattern

%%%%%%%%%%%%%
% equations:
% Signal to noise ratio (achieving optimality)
% Observed signal model
%%%%%%%%%

\begin{figure}
  \includegraphics[width=1.0\textwidth]{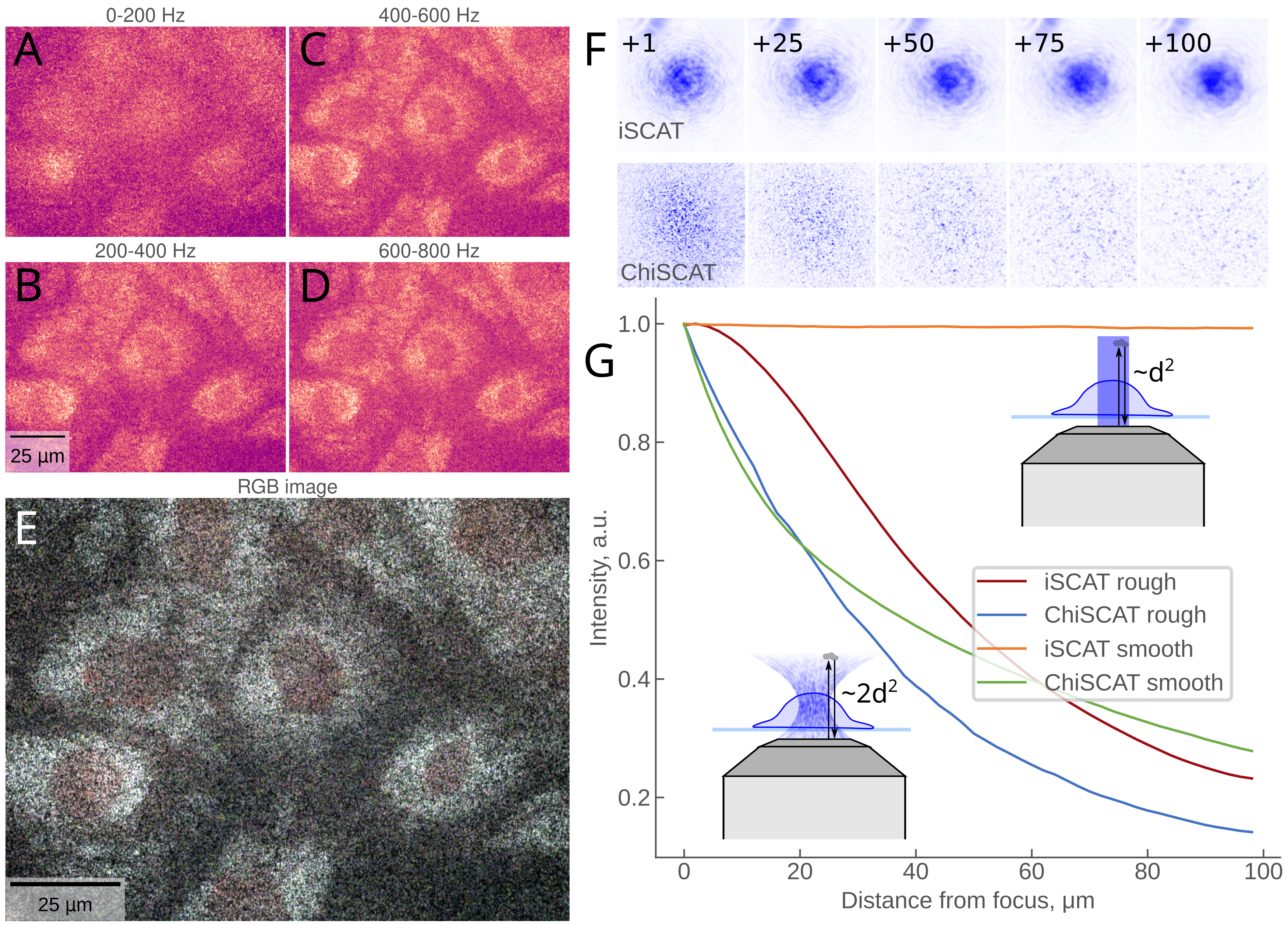}
  \caption{Frequency-resolved cellular dynamics imaging by ChiSCAT imaging. A-D: rms activity in a band-pass-filtered ChiSCAT recording. E: visualisation as an RGB image (red channel = B, green channel = C, blue channel = D), a region with reduced high-frequency fluctuations is visible in the center of the cell, presumably the nucleus and/or reticulum. F: Measurement of depth-selectivity. % Images were captured while varying the distance between objective and coverslide.
  The data presented are camera images of a coverslide moving out of focus for iSCAT and ChiSCAT illumination. The number in the corner denotes the distance from the focal plane in $\mu$m. 
  G: observed signal intensity as a function of distance from focus, obtained by integrating across a region of interest in F. 
  We compare a smooth sample (cover slide, traces "iSCAT smooth", "ChiSCAT smooth") and a rough sample (polymer microspheres, "iSCAT rough", "ChiSCAT rough"). In all cases, the ChiSCAT signal intensity drops significantly faster then iSCAT. }

  % RGB images: exp_id = 287
  %To check that collected speckle signal indeed bears information about cell, exploratory data analysis was conducted. Assuming validity of signal model, shown in the Figure \ref{fig:setup}, D, high-frequency components of the signal should only be given by dynamic cell membrane fluctuations and noise. To extract this information, Fourier transform was run on the sample of the speckle video, and energies in the different frequency bands were integrated to get average cell activity in the given frequency region (Figure \ref{fig:rgb_image}). Since different parts of the cell have different stiffness, they should exhibit different frequential properties. To effectively compare them, 3 frequencial bands (200-400 Hz, 400-600 Hz, and 600-800 Hz) were stacked and visualized as individual channels of RGB image, shown in the Figure \ref{fig:rgb_image}. 
  \label{fig:rgb_and_intensity}
\end{figure}

The ChiSCAT recording is interferometrically sensitive to motion, and meaningful images can be reconstructed from the chaotic speckle pattern by data processing. As a simple example, temporal band-pass filtering of the video stream reveals an easily recognizable image of individual cells. Fluctuations at high ($>300$ Hz) frequencies are markedly reduced in or around the nucleus, possibly due to restricted diffusion in the nucleus and reticulum, confirming recent observations by refractive index tomography \cite{hugonnet23}. ChiSCAT is depth-selective. The signal observed in a region of interest arises mainly from a range of depths comparable to the size to the region, because ChiSCAT, in contrast to existing schemes \cite{kukura2009high,popescu2019quantitative} employs high-NA illumination. We demonstrate this by measuring the suppression of light scattered from a movable out-of-focus plane (Fig. \ref{fig:rgb_and_intensity}, F + G).

\section{Data processing}

\subsection{Preliminary analysis}

\begin{figure}
  \includegraphics[width=0.5\textwidth]{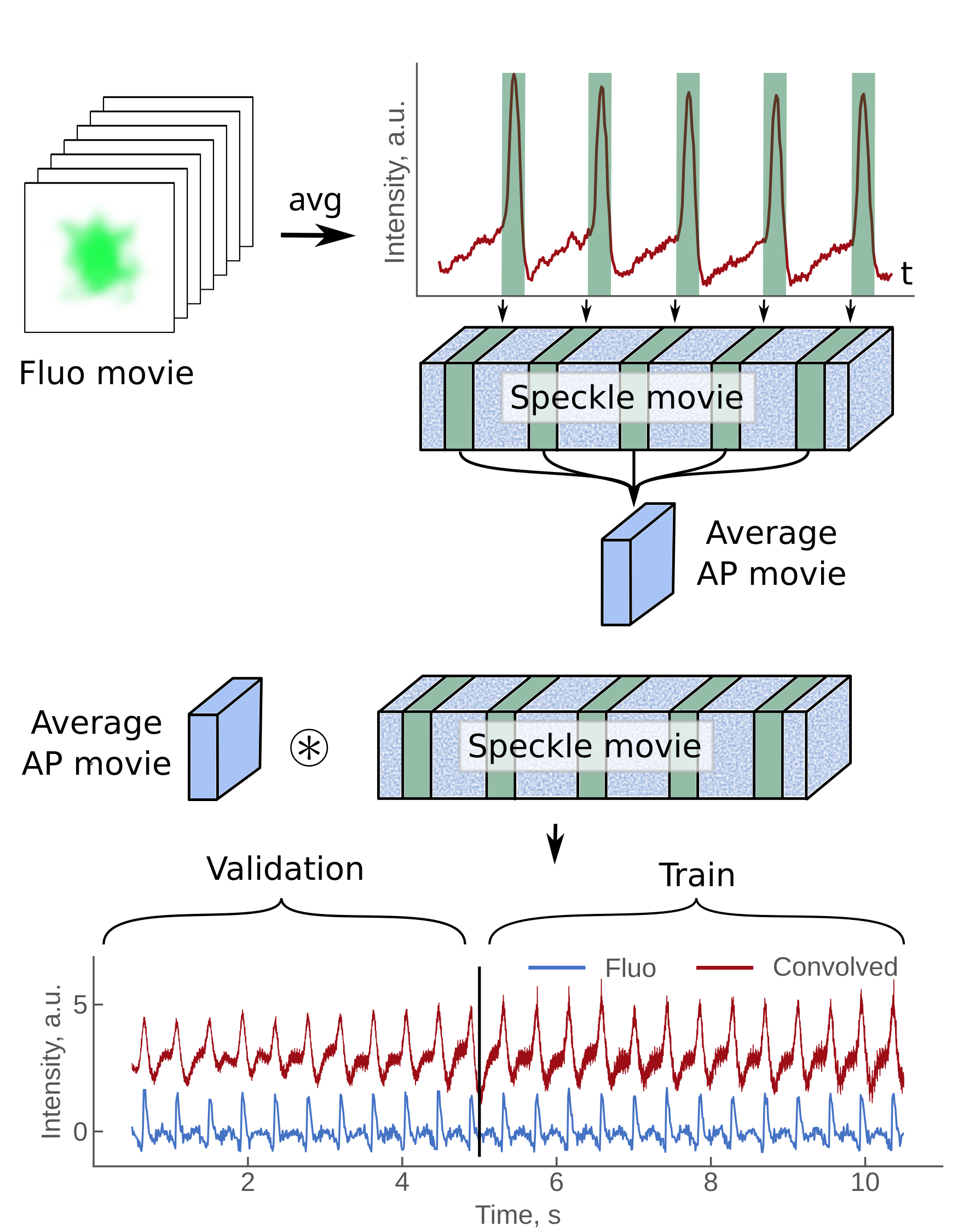}
  \caption{Supervised match filtering approach. Ground truth AP is extracted from the training subset of the fluorescence movie to calculate average spike-triggered speckle movie. The latter is then used as convolution filter on the rest of the speckle signal, allowing to extract AP from the validation part of the movie.}
  \label{fig:algo_naive}
\end{figure}

\subsection{Supervised learning}
We now turn to the main goal of this work: label-free detection of action potentials from a visually non-interpretable ChiSCAT recording. In mathematical terms, the goal is finding $a(t)$, the time-varying amplitude of AP activity with maximum signal-to-noise ratio. 

It is known that the optimal linear solution to this problem is matched filtering \cite{north1963analysis}. This scheme provides an estimate $a_{\text{MF}}(t)$ of $a(t)$ by convolving the signal (in our case a ChiSCAT movie $I_{\text{all}}$) with a time-reversed elementary template $a_{\text{el}}$
$$
a_{\text{MF}} = I_{\text{all}} \circledast_t a_{\text{el}} = \sum_{x,y,t} I_{\text{all}}(x,y,t) I_{\text{el}}(x,y,-t)
$$
In our case, $I_{\text{el}}= I_{\text{AP}}\otimes a_{\text{el}}(t)$ can be thought of as a noise-free ChiSCAT movie of a perfect single action potential. Crucially, neither $I_{\text{AP}}$ nor $a_{\text{el}}$ is known in advance. 

% \todo{fix math, use different styles for matrices, tensors and vectors}

One way to obtain $I_{\text{AP}}$ is supervised learning by spike-triggered averaging of the ChiSCAT signal, which we demonstrate in Figure \ref{fig:algo_naive}. We first extract a fluorescence trace by spatially averaging the fluorescence movie. Fluorescence peaks, corresponding to individual APs, are detected and used to extract slices from the ChiSCAT movie (only training subset), which are then averaged to produce an estimation of $I_{\text{AP}}$. The original speckle movie is then convolved with this extracted matched filter to produce the recovered signal $a_{\text{MF}}(t)$.

% We assume that the observed signal is a mixture of cell speckle pattern, weighted by temporal pattern, equal to AP signal (Figure \ref{fig:setup}, D) plus noise. Under these assumptions, the optimal matched filter is simply the cell speckle pattern, weighted by one period of AP signal, which can be recovered by simply averaging spike-triggered videos. 
% The approach is shown in the Figure \ref{fig:algo_naive}. 

As visible from Figure \ref{fig:algo_naive}, this approach is indeed able to recover individual APs, however it suffers from two problems: firstly, it requires a known ground truth signal for training (fluorescence in our case), which undermines the concept of label-free microscopy. Secondly, this approach is only optimal under assumption of static background (static speckle $I_{\text{static}}$ in the Figure \ref{fig:setup}, D) and a perfectly repeating dynamic pattern  $I_{\text{AP}}$, while in reality both slowly drift on a scale of 2-3 seconds due to slow motion of the cell. This makes the spike-triggered average incoherent and reduces the signal. In an experiment on 64 regions of cells, supervised learning was successful in 36\% of the regions. 

\subsection{Unsupervised learning}
The most central discovery of this work is an unsupervised learning scheme that can efficiently recover repeating patterns of motion without requiring ground truth information. It instead exploits the {\em a priori} information that APs are recurring events and that the approximate temporal amplitude pattern $a_{\text{el}}$ of an AP is known. 

\begin{figure}
  \includegraphics[width=1.0\textwidth]{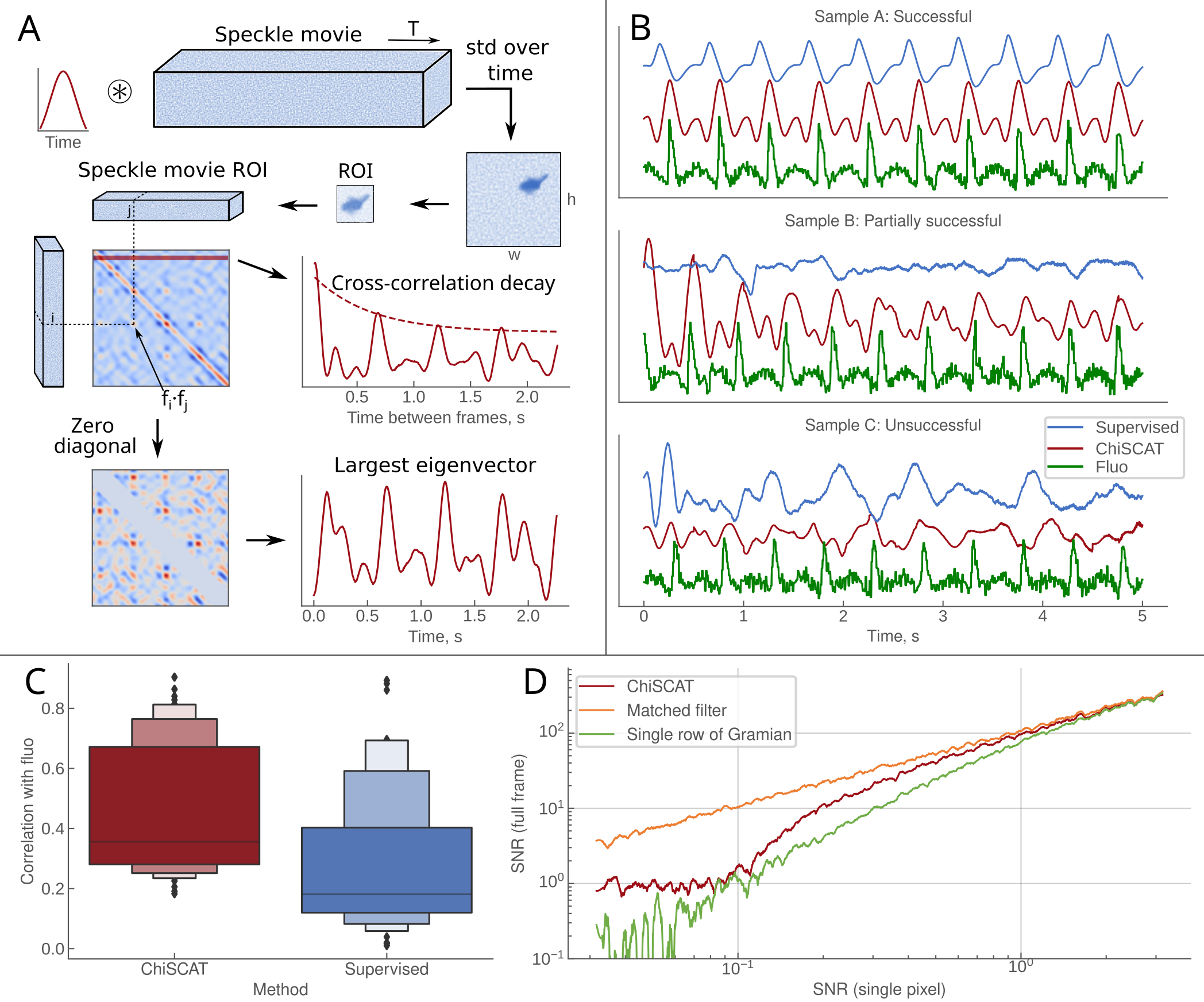}
  \caption{Unsupervised reconstruction of action potentials. (A) Signal processing pipeline. The ChiSCAT movie is matched-filtered by convolution with a temporal template of the action potential. Recurring events are identified by computing the Gramian matrix $G_{ij}=f_i \dot f_j$ for each pair of frames $f_i, f_j$. Recurring events appear as off-diagonal peaks in this matrix, and a column or row crossing one of the peaks ("Cross-correlation trace") is a time trace of the AP activity. A full-length AP trace ("largest eigenvector") is obtained by eigendecomposition of $G$. (B) reconstructed AP activity for different samples of cells. (C) Distribution of correlation coefficient between extracted signal and ground truth for different samples of cells. ChiSCAT restores the AP signal in the majority of cases. (D) Simulated SNR of the extracted signal for different algorithms as a function of SNR in a single pixel ($N_{\text{px}}=100^2$, 9 APs in the simulated data). The full ChiSCAT algorithm demonstrates an intermediate performance between matched filtering with a perfect template and simple extraction of off-diagonal peaks in the Gramian.
  % exp_ids: Sample A: 289, B: 291, C: 290
  % (C) Optimality proof for the unsupervised algorithm. Considering every frame as an $N_{\text{px}}$-dimensional vector ($N_{\text{px}}$: number of pixels) the dot product between two AP frames will be dominated by the parallel AP signal component of the vectors, even in presence of a random added noise vector $\vec N$, provided that $|\vec N| < |\vec S|$.   
  }
  \label{fig:algo_chiscat}
\end{figure}

The full algorithm is visualized in the Figure \ref{fig:algo_chiscat} (A). It is based on a combination of temporal matched filtering and motif discovery. %better term?
The algorithm begins by convolving the raw ChiSCAT movie along the time axis with a 1D model of the signal $a_{\text{el}}$. For simplicity, a zero-mean Gaussian function was chosen, the width $\sigma$ of which was adapted to the typical timescale of an AP. 
%as the model because it has just one parameter, $\sigma$, facilitating easy tuning. In our case, the $\sigma$ was chosen based on typical timescale of AP, known from fluorescence measurements, however it worth noting that the algorithm is insensitive to this parameter and similar results were achieved even by scaling $\sigma$ by a factor of 5

%We propose another approach, free from limitations of the naive one. The approach is based on the idea that two frames with AP happening should be more similar to each other then two frames containing only noise. The full algorithm is visualized in the Figure \ref{fig:algo_chiscat} (A). 

The resulting convolved movie was subsequently cropped to a region of interest (ROI), identified from pixels with high temporal variability. This is accomplished by computing the standard deviation along the time axis, followed by Gaussian smoothing and thresholding.    
%Next, regions of interest (ROIs) are identified within the field of view (FOV) by exploiting the higher variability in pixels containing cell signal. 

We then search for recurring patterns in the resulting movie by calculating the dot product between every possible pair of frames, forming a Gramian matrix. A recurring events, i.e. a set of two similar frames occurring some time apart, will produce an off-diagonal peak in this matrix because the frames are strongly correlated with each other, whereas each of them is uncorrelated with the noise. This approach to identifying recurring events is reminiscent of a recurrence plot \cite{kamphorst1987recurrence}, only using a dot product rather than a difference as a metric for similarity. APs can be detected as off-diagonal peaks of the Gramian matrix, and a column or row crossing one of the peaks is a time trace of AP activity. 

Due to the slow fluctuations of the cell, the cross-correlation between two frames both containing an AP decreases with increasing temporal distance between them, and vanishes for timescales larger than few seconds. To reconstruct a non-decaying AP trace over the ChiSCAT movie, we use a rank-one approximation, that is we find a vector $a_{\text{rec}}$, such that its outer product with itself $a_{\text{rec}} a_{\text{rec}}^\intercal$ approximates the Gramian matrix. 
One can imagine this as finding a checkerboard pattern in the matrix which best approximates the experimentally observed Gramian.
Technically, this is achieved by finding the largest eigenvector of the Gramian. Before the eigendecomposition, the diagonal of the Gramian matrix is zeroed out to reduce the influence of random non-recurrent events. Importantly, the low-rank approximation aggregates information from multiple action potentials and hence improves the signal-to-noise ratio over what could be obtained in a single correlation of one pair of frames. 

Applying the algorithm to a set of cell regions (Fig.~\ref{fig:algo_chiscat} B), we find that it indeed recovers action potentials, even for regions where supervised learning has failed. Overall, detection succeeds in 71\% of the regions (Fig.~\ref{fig:algo_chiscat} C), which is a marked improvement over supervised detection. 

We can confirm this improvement by a simple numerical model (Fig.~\ref{fig:algo_chiscat} D). 
It models the frames $\vec{f}_i$ of the speckle movie as random Gaussian noise with an additional fixed zero-mean random pattern $\vec{p}$ periodically superimposed onto some selected frames to model the signal. This effectively models action potentials as a train of delta functions in time. 
From this synthetic dataset we compute three time traces: 1) The time series $r_i = \vec{f}_i \cdot \vec{p}$, effectively modeling spatial matched filtering with a perfect template $\vec{p}$. 2) The time series $r_i = \vec{f}_s \cdot \vec{f}_i$, where $\vec{f}_s$ is a signal frame containing both the signal pattern $\vec{p}$ and Gaussian noise, effectively modeling a single column or row of the Gramian matrix in Fig.~\ref{fig:algo_chiscat}. 3) The time series obtained from the full ChiSCAT algorithm, i.e. all steps of Fig.~\ref{fig:algo_chiscat} A. For all series, we compare the height of a signal peak with the rms noise of a signal-free region. 
As expected, spatial matched filtering performs best, and can be fully matched by the other schemes only if a signal-to-noise ratio larger than one is already obtained in a single-pixel time trace. Unsupervised detection correlates the movie with a signal frame instead of the perfect pattern $\vec{p}$. For weaker signals, this signal frame is dominated by noise, which leads to inferior performance. A simple analytical estimate (SI) suggests that this overhead is a factor of $\sqrt[4]{N_{\text{px}}}$ in SNR, where $N_{\text{px}}$ is the number of pixels in a frame. Due to the slow ($ \sqrt[4]{\;} $) scaling, this overhead is not excessive, reducing the SNR by one order of magnitude for a $100 \times 100$ pixel movie. The full ChiSCAT algorithm recovers for part of this loss, because it searches for an extended pattern of cross-peaks in the Gramian matrix and hence aggregates information from multiple action potentials. Over a wide range of single-pixel signal-to-noise ratios, the resulting performance approaches matched filtering. 

\section{Results and discussion}

 % * chiscat maximally sensitive to motion, experimentally simple, rejection of vibrations, wavelength can be chosen. 
 % * unsupervised learning can recover APs without a priori knowledge
 % * (real cells)
 % * outperforms supervised scheme 

In summary, we have demonstrated an unsupervised learning algorithm that can recover action potentials from a microscopy movie, even from a chaotic speckle pattern, without {\em a priori} information on the timing and the exact spatial pattern of an AP. With this tool at hand, we detect action potentials with interferometric sensitivity in an experimentally simple microscopy scheme. Combining iSCAT microscopy with speckle illumination, makes it interferometrically sensitive to motion. It harnesses the higher sensitivity that can be achieved in reflective schemes and features common-path rejection of vibrations and a freely tunable illumination wavelength. 

Our results also demonstrate that living cells are interferometrically stable for at least a few seconds and that APs repeating within this time indeed trigger the same interferometric pattern. This opens a window where algorithms can be devised to search for repeating signals rather than a known pattern. 

 % * discussion of 4sqrt overhead
 %    - fewer pixels better
 %    - but: need to distinguish motion patterns, compromise

The signal to noise ratio of our algorithm is inferior to algorithms searching for a known spatial pattern, and scales as
%The amount of photons, needed to achieve SNR=1 scales as 
$\sqrt[4]{N_\text{px}}$, i.e. for a realistic setup having $100 \times 100$ pixels ChiSCAT will be 10 times less sensitive than matched filtering. The theory predicts that lowering the number of pixels will improve the SNR, achieving optimality in the limiting case of $N_\text{px}=1$. However, this would make distinguishing between an AP signal and a random fluctuations impossible, since both of them will exhibit exactly same spatial pattern (consisting of 1 pixel). This trade-off leads to the existence of an optimal number of pixels for the given experiment, which will need more investigation in future work.

 % * move to neurons -> smaller motion, but more spikes. 

Recovery of signal with ChiSCAT algorithm depends not only on the number of pixels, but also on the density of spikes: the low rank approximation minimizes the sum of squares of residuals, i.e. the algorithm will converge to a component which dominates the Gramian matrix. Increasing the number of spikes increases the number of off-diagonal peaks in the Gramian matrix quadratically, helping the algorithm to recover the real AP signal. Thus, the theory predicts that the ChiSCAT algorithm can recover even weaker signals if cells spike more often, e.g. in primary neurons.

 % * works on any microscopy technique, could identify patterns in imaging microscopy
 % * could detect any pattern, not only plasma membrane motion. E.g. Ca release, birefringence change, lateral motion (El-Hady theory)
 % * works on speckles = works in scattering tissue -> living brain

While our demonstration is based on a speckle-based variant of iSCAT microscopy, the algorithm is agnostic against the imaging scheme employed, and could also operate on visually interpretable imaging datasets obtained by e.g. quantitative phase microscopy, ROCS, refractive index tomography or coherence tomography \cite{popescu2019quantitative,junger2016fast, charriere2006cell,optical_coherence_tomography}.
Importantly, it could identify any repeating motion correlating with action potentials, not only motion of the plasma membrane. It could thus discover additional intrinsic signals, such as motion of organelles in response to Ca release, or a lateral motion in the axoplasm which is predicted by a non-standard theory of action potentials~\cite{elhady2015mechanical}. 
Still, it is notable that our algorithm can even operate on speckle patterns. This paves the way to detecting action potentials even in scattering tissue, including the living brain.

%set of discrete repeating signal frames, superimposed to random (Gaussian) noise.
%A signal frame $\vec{s}_i$ is a $m\times n$ picture with (Gaussian) random values at every pixel, which however is the same frame for all signal frames. 
%The model thus assumes that the signal occurs only within a single frame, which neglects the extended temporal shape of the action potential. 

% Since that approximately corresponds to the size of ROI in the experiments, ChiSCAT approach achieves approximately 10x lower SNR than optimal approach, which is still affordable since theoretical calculations above show that in ideal case we expect SNR of 100.

% Compressing the cell into a lower number of pixels thus leads to a higher SNR. However, pixel size must not be significantly larger than diffraction limit, otherwise points havi

% more pixels can discern different motion patterns and can separate membrane fluctuation from e.g. nucleus motion. 
% a sum of , 

% Fig C: biologists' figure
% Fig D: simulation resutlsunsupervised

% The unsupervised approach is able to detect APs in xxx percent of the cell regions studied (Fig. \ref{fig:algo_chisat}). 

\section{Methods}

All the experiments have been performed with custom-built microscope (see SI for details). The commercially available iPS cell line SC950A-1 (Biocat) was cultivated in StemMACS iPS-Brew (Miltenyi Biotech). Once the cells had reached 80\% confluence, they were split as aggregates using ReLeSR (Stemcell technologies) according to the manufacturer's instructions and seeded at a cell density of 20,000 cells/cm2. 
For the selection of cardiomyocytes, clones with the $\alpha$MHC promoter \cite{klug1996genetically} and for the selection of pacemaker cells, double transfected clones with TBX3 and $\alpha$MHC promoter \cite{jung2014programming} were generated. Four days before the start of differentiation, 20000 cells/cm2 were seeded into one well of a 6 well plate in StemMACS iPS-Brew. Differentiation protocol was applied. After end of differentiation, cells were dissociated using the Multi Tissues Dissociation Kit 3 (Miltenyi Biotech) and seeded as single cells for further analyses. See SI for detailed information about methods.

% Details about the cells

\begin{acknowledgement}
The authors wish to thank Marten Möller for expert technical assistance.
This work has been supported by the Deutsche Forschungsgemeinschaft (DFG) SFB 1477 “LightMatter Interactions at Interfaces,” Project No. 441234705. S.K., C.R. and R.D. received funding from the EU Structural Fund (ESF/14- BM-A55-0024/18) and the Federal Ministry for Economic Affairs and Climate Action (16KN083635). In addition, S.K. is supported by the DFG (VA1797/1-1). Moreover, R.D. is supported by the DFG (DA1296/6-1 and GRK 2901 Sylobio), the DAMP foundation, the German Heart Foundation (F/01/12), the BMBF (VIP+00240) and the BMEL (FKZ: 281A819B21). 
\end{acknowledgement}

\begin{suppinfo}

% This will usually read something like: ``Experimental procedures and
% characterization data for all new compounds. The class will
% automatically add a sentence pointing to the information on-line:

Optical setup description, details about numerical and analytical estimation of ChiSCAT algorithm performance.

\end{suppinfo}

\bibliography{chiscat}

% \todo{Send article to prof David}
% \todo{upload minified data + jupyter to a repository and create a link on Kaggle}
% \todo{add IDs of experiments as comments}

\end{document}

\section{Setup details}

Setup (Fig. 1 (A) in the main text) consists of illumination laser (473 nm, CNI Laser), which is used for both fluorescence excitation and speckle imaging. Firstly, the laser beam is expanded (to fill the objective aperture) and illuminates holographic diffuser (Luminit, 5° diffisuing angle). Holographic diffuser possess well defined angle of diffusion, which minimizes amount of light, not captured by lens. Lens 1 (Thorlabs LB1607) is a simple relay lens which projects illuminated diffuser image onto objective backfocal plane (BFP), thus creating Fourier-transformed random field (which is still a random field) in the cell plane. An important aspect of projecting to BFP, is that by changing diffuser angle, it is possible adjust the FOV. Reflected blue laser and excited fluorescence are extracted with 50:50 beamsplitter (Thorlabs BS031), and split according to wavelength with dichroic mirror (DM, Thorlabs MD499). The fluorescence channel is additionally filtered with a bandpass filter corresponding to the dye emission (Thorlabs, MF530-43) and focused into EMCCD camera sensor (Princeton Instruments ProEM HS 512x512) with the Lens 2 (Thorlabs LA1708-A). Speckle channel is imaged onto CMOS camera (KronTech Chronos 2.1-HD) with the Lens 3 (Thorlabs AC-508-150).

All experiments were performed at 2000 FPS CMOS camera at resolution 640x480 and 100 FPS of EMCCD camera, i.e. every fluorescence frame was recorded per 20 speckle frames. Laser power was set to 35 mW.

% Synchronization details??

\section{Correlation computation}

% Correlation between reconstructed signal ($s$) and fluorescence trace ($f$), presented in the Fig. 4 (C) in the main text was calculated by 
% - Computing cross-correlation between fluorescence trace and chunk of reconstructed signal and taking its maximum value
% - Sliding the chunk along the reconstructed trace 
% - Taking maximum value for all chunks

The correlation between the reconstructed signal and the fluorescence trace, as depicted in Figure 4 (C) in the main text, was determined as follows: Firstly, the cross-correlation between the fluorescence trace and a chunk of the reconstructed signal was computed, and its maximum value was taken. Then, this process was repeated by sliding the chunk along the reconstructed trace, followed by taking globally maximum correlation value. This is necessary to guarantee that correlation is not biased by small shifts between traces or by random events in the trace. The correlation was only computed for subset of experimental data (69 traces in total) where fluorescence measurement was reliable enough.

\section{Modeling of the signal-to-noise-ratio (SNR) of ChiSCAT}
\subsection{Toy model for SNR analysis (Fig.~4 D main manuscript)}
To estimate the signal-to-noise ratio that can be obtained by the ChiSCAT algorithm, we employ a simple toy model of the signal obtained after the initial matched filtering step. We model the matched-filtered speckle movie as a sum of photonic shot noise, represented by a series of frames where every pixel is drawn from a normal distribution with standard deviation $\sigma = 1$, and synthetic action potentials, modeled by a repeating signal-pattern $\vec{p}$ that is added to specific selected frames. The signal pattern $\vec p$ is equally drawn from a normal distribution, which however has a different standard deviation $\sigma=\text{SNR}_{\text{1px}}$ and is moreover identical  for all signal frames. Note that action potentials are modeled as delta-function events, which is an approximation. We will denote the frames of the resulting synthetic data movie as $\vec f_i$. 

\begin{figure}
  % As well as the standard float types \texttt{table}\\
  % and \texttt{figure}, the class also recognises\\
  % \texttt{scheme}, \texttt{chart} and \texttt{graph}.
  \includegraphics[width=1.0\textwidth]{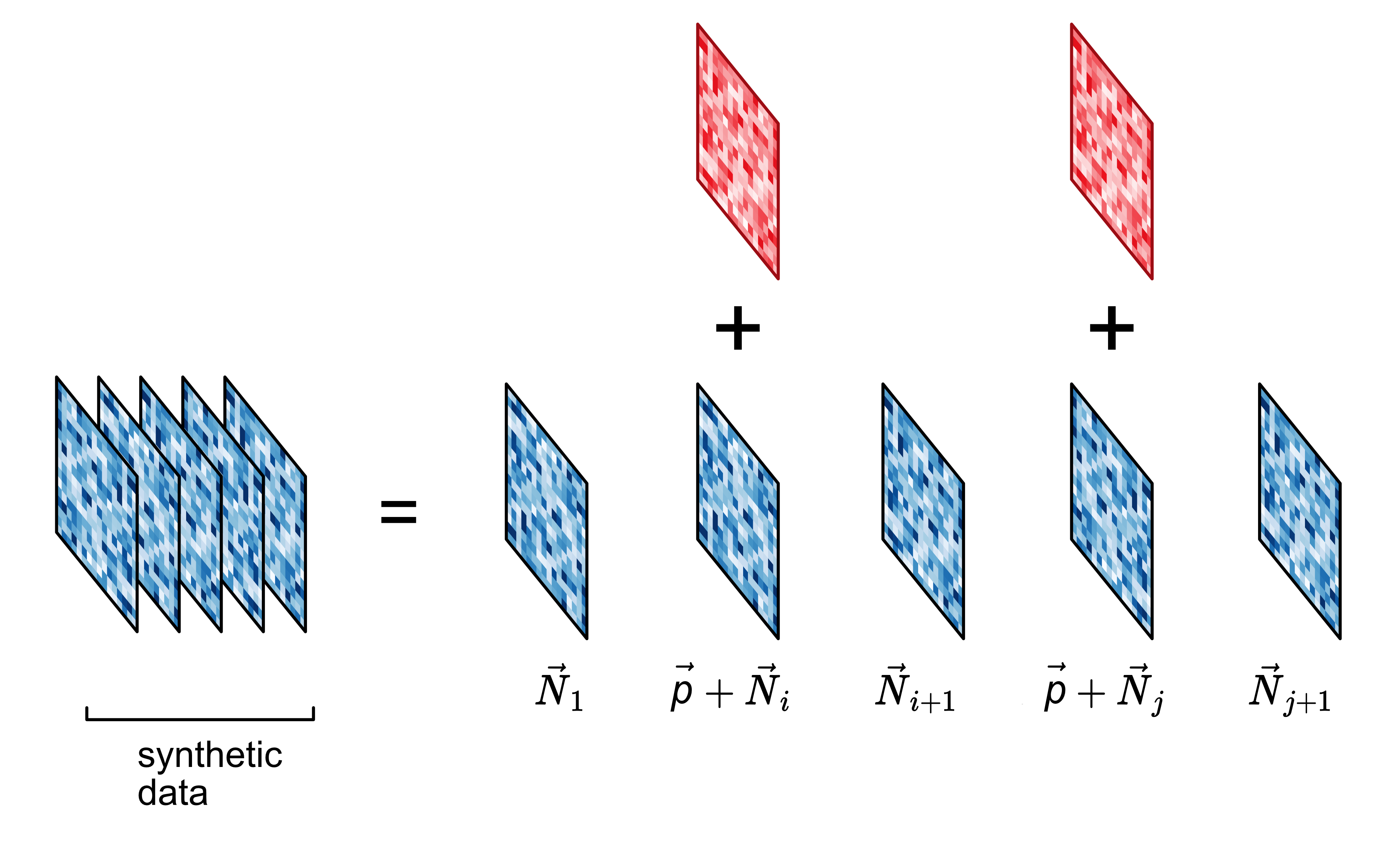}
  \caption{Toy model of data for the signal-to-noise-analysis (Fig.~4 D main text). A matched-filtered speckle movie is approximated as a series of noise frames $\vec N_k$, where every pixel is drawn from a normal distribution with variance $\sigma=1$. A repeating signal-pattern $\vec p$ is added to specific selected frames (denoted $i$ and $j$) to model action potentials. This signal pattern $\vec p$ is drawn from a normal distribution with $\sigma=\text{SNR}_{\text{1px}}$ but is identical  for all signal frames. 
  Note that pixel values can be negative because we model the movie after an initial step of matched filtering. }
  \label{fig:setup}
\end{figure}

We analyze this synthetic data in three ways: 
\begin{enumerate}
    \item by simulated spatial matched filtering with the perfect signal template $\vec p$, i.e. computing the time series $r_i = \vec f_i \cdot \vec p$.
    \item by computing a single column of the Gramian matrix at a signal frame $\vec f_k$, i.e. computing the time series $r_i = \vec f_i \cdot \vec f_k$. 
    \item by running the full ChiSCAT algorithm, i.e. computing the Gramian matrix $\vec f_i \cdot \vec f_k$, erasing its diagonal and computing the largest eigenvector as a rank-one approximation. 
\end{enumerate} 
For each of these time series, we extract the signal strength by calculating $\vec r_j$ at a signal frame $j$, and we extract the background noise by numerically calculating the standard deviation of $\vec r_i$ in a signal-free time window. 
\subsection{Analytical estimate of SNR}
One explicit design feature of ChiSCAT is to only use minimal {\em a priori} information. In particular, we do not assume knowledge of the perfect signal pattern $\vec p$. This feature comes at a price, which is a performance overhead compared to algorithms like matched filtering, which assume knowledge of the signal pattern. Quantitatively, this loss can be estimated by comparing the SNR of a single off-diagonal peak in the Gramian matrix to the SNR obtained in perfect matched filtering. 
The strength of an off-diagonal Gramian peak is the scalar product between two signal frames $i$ and $j$
$$
\vec f_i \cdot \vec f_j = (\vec p + \vec N_i)\cdot (\vec p  + \vec N_j) \approx \vec p \cdot \vec p + \vec N_i \cdot \vec p + \vec N_j \cdot \vec p + \vec N_i \cdot \vec N_j 
$$
where the first term $\vec p \cdot \vec p$ constitutes the signal, and the latter three terms constitute noise. For the most relevant parameter range $\text{SNR}_\text{1px}<1$, the dominant noise term is the last term $\vec N_i \cdot \vec N_j$.
The strength of a signal peak in matched filtering is the scalar product of a signal frame with the perfect pattern $\vec p$
$$
\vec f_i \cdot \vec p = (\vec p + \vec N_i)\cdot \vec p \approx \vec p \cdot \vec p + \vec N_i \cdot \vec p.
$$
Again, the first term constitutes the signal, the latter term the noise. \par

We now estimate the strength of all individual terms appearing in the expressions above. \par 
\begin{itemize}
    \item $\langle \vec p \cdot \vec p \rangle = \text{SNR}_\text{1px}^2 {N_{\text{px}}}$, because for every pixel $p_m$ of $\vec p$, the product $p_m^2$ is a random variable drawn from a chi-square distribution with mean value $\text{SNR}_\text{1px}^2$.
    \item $\langle \vec p \cdot \vec N_i\rangle = 0, \sqrt{{\left\langle \left(\vec p \cdot \vec N_i\right)^2\right\rangle }} = \text{SNR}_\text{1px} \sqrt{N_{\text{px}}}$. For every pixel the product of $\vec p$ and $N_i$ is the product of two uncorrelated Gaussian random variables, with variances $\sigma_p^2 = \text{SNR}_\text{1px}^2$ and $\sigma_i^2 = 1$. The variance of the product is known to be $\sigma_{\text{product}}^2 = \sigma_p^2\sigma_j^2 = \text{SNR}_\text{1px}^2$.
    \item $\langle \vec N_i \cdot \vec N_j \rangle = 0, \sqrt{{\left\langle \left( \vec N_i \cdot \vec N_j\right)^2 \right\rangle}} = \sqrt{N_{\text{px}}}$. Similarly, for every pixel the product of $N_j$ and $N_i$ is the product of two Gaussian random variables, with variances $\sigma_i^2 = 1$ and $\sigma_j^2 = 1$. The variance of the product is thus $\sigma_{\text{product}}^2 = \sigma_i^2\sigma_j^2 = 1$.
\end{itemize}
With these expressions, the condition $\text{SNR} > 1$ for matched filtering translates into the equation 
\begin{eqnarray}
\vec p\cdot \vec p &>& \sqrt{{\left\langle \vec p \cdot \vec N_i\right\rangle }^2}\\
\text{SNR}_\text{1px}  &>& \frac {1}{\sqrt{{N_{\text{px}}}}}
\end{eqnarray}
consistent with the intuition that averaging over $N_{\text{px}}$ pixels improves the SNR by $\sqrt{{N_{\text{px}}}}$.\par
In contrast, the condition $\text{SNR} > 1$ for the initial step of ChiSCAT reads
\begin{eqnarray}
% \vec p\cdot \vec p &>& \sqrt{{\left\langle \left(\vec N_i \cdot \vec N_j \right)^2\right\rangle}}\\
\text{SNR}_\text{1px}^2 \sqrt{{N_{\text{px}}}} &>& 1 \\
\text{SNR}_\text{1px}  &>& \frac{1}{\sqrt{{N_{\text{px}}}}} \cdot \sqrt[4]{{N_{\text{px}}}}\\
\end{eqnarray}
Compared to matched filtering, there is an overhead of $\sqrt[4]{{N_{\text{px}}}}$ in terms of SNR, which can be partially recovered in the final step of ChiSCAT by merging information from multiple action potentials.

\section{Methods}

Differentiation protocols for human iPS cells to generate CM mainly result in a mixtures of the three subtypes, with ventricular CMs predominating and atrial and pacemaker cells underrepresented. We have focused on the function of cardiac transcription factors that regulate both cardiogenesis during embryonic development and the cardiac differentiation of pluripotent stem cells. The principle of cardiomyocyte subtype-specific stem cell programming by overexpression of transcription factors (forward programing) was used to program pluripotent stem cells into previously unobserved aggregates of pacemaker cells \cite{jung2014programming,rimmbach2015generation,yavari2017mammalian}. Electrophysiological parameters like the peak voltages of an action potential, its duration, or its delay with respect to Calcium release and contraction, are highly important observables to characterize the cell type and function, but currently require electrophysiology or invasive labels to be observed.

\subsection{Cultivation of human induced pluripotent cells}

The commercially available iPS cell line SC950A-1 (Biocat) was cultivated in StemMACS iPS-Brew (Miltenyi Biotech). Cell culture vessels were coated with Laminin 521 (Biolamina) according to the manufacturer's instructions. Once the cells had reached 80\% confluence, they were split as aggregates using ReLeSR (Stemcell technologies) according to the manufacturer's instructions and seeded at a cell density of 20,000 cells/cm2. Medium was changed daily.

\subsection{Generation of stable IPS clones}

For the selection of cardiomyocytes, clones with the $\alpha$MHC promoter (Klug et al. 1996) and for the selection of pacemaker cells, double transfected clones with TBX3 and $\alpha$MHC promoter \cite{jung2014programming,rimmbach2015generation} were generated as follows:

One hour before nucleofection, the medium was removed and fresh StemMACS iPS-Brew containing 10 µM Rock inhibitor and 10\% Nate (Invivogen) was added.

For nucleofection, the cells were separated in the following way: Medium was removed and cells were washed with PBS. The cells were then detached with Accutase (37°C, 5 min) and separated. After the addition of DMEM, the cells were removed from the culture vessel and transferred to a centrifuge tube. The cell suspension was centrifuged (5 min, 300xg) and the pellet was resuspended in IPS Brew with 10 µM Y-27632.

For nucleofection, 100000 cells were taken up in buffer P3 with supplement 1 (Lonza) and 5 µg endotoxin-free plasmid DNA ($\alpha$MHC promoter or TBX3) was added.

Transfection was performed in the 4D Nucleofector (Lonza) using the CB150 programme. The cells were seeded at a density of 65,000 cells/cm2 in StemMACS iPS-Brew with 10 \% CloneR (Stem cell technologies) and cultured for 24 h. The cells were then selected with 100 µg/ml hygromycin B (cardiomyocytes; $\alpha$MHC promoter). Subsequently, selection was performed with 100 µg/ml hygromycin B (cardiomyocytes; $\alpha$MHC promoter) or 100 µg/ml hygromycin B and 10 µg/ml blasticidin (pacemaker cells; TBX3, $\alpha$MHC promoter). Individual clones were harvested and propagated.

\subsection{Differentiation}

In all steps the medium was changed daily, before each change to a different medium the cells were rinsed twice with PBS. Four days before the start of differentiation, 20000 cells/cm2 were seeded into one well of a 6 well plate in StemMACS iPS-Brew.

On day 1 of differentiation, the medium was changed to RPMI with B27 without insulin (Thermo Fisher), 200 µM ascorbic acid, 5 ng/µl FGF2, 1µM CHIR99021, 5 ng/µl BMP4 and 9 ng/µl Activin A. From day 4, the cells were differentiated in RPMI with B27 with insulin, 200 µM ascorbic acid and 5 µM IWP2. On day 10, a switch was made to RPMI with B27 with insulin and 200 µM ascorbic acid. The selection of cardiomyocytes (beginning on day 14) was performed in RPMI with B27 with insulin, 200 µM ascorbic acid with 200 µg/ml G418. On day 18, cardiomyocytes were detached using the Cardiomyocyte Isolation Kit (Stemcell technologies), seeded into one well of a 24 well and cultured in Cardiomyocyte Maintenace Medium (Stemcell Technologies). Once these had formed a stable contracting syncytium, they were dissociated using the Multi Tissues Dissociation Kit 3 (Miltenyi Biotech) and seeded as single cells for further analyses.

\bibliography{chiscat}